# HERMES: qualification of High pErformance pRogrammable Microprocessor and dEvelopment of Software ecosystem

Nadia Ibellaatti*, Edouard Lepape*, Alp Kilic*, Kaya Akyel*, Kassem Chouayakh*, Fabrizio Ferrandi†, Claudio Barone†, Serena Curzel†, Michele Fiorito†, Giovanni Gozzi†, Miguel Masmano‡, Ana Risquez Navarro‡, Manuel Muñoz‡, Vicente Nicolau Gallego‡, Patricia Lopez Cueva§, Jean-noel Letrillard¶, Franck Wartel**

*NanoXplore, France, †Politecnico di Milano, Italy, ‡Fent Innovative Software Solutions – FentISS, Spain, §Thales Alenia Space SAS, France, ¶STMicroelectronics Grenoble 2 SAS – STGNB 2, France, **Airbus Defence And Space SAS, France

*Abstract* — European efforts to boost competitiveness in the sector of space services promote the research and development of advanced software and hardware solutions. The EU-funded HERMES project contributes to the effort by qualifying radiation-hardened, high-performance programmable microprocessors, and by developing a software ecosystem that facilitates the deployment of complex applications on such platforms. The main objectives of the project include reaching a technology readiness level of 6 (i.e., validated and demonstrated in relevant environment) for the rad-hard NG-ULTRA FPGA with its ceramic hermetic package CGA 1752, developed within projects of the European Space Agency, French National Centre for Space Studies and the European Union. An equally important share of the project is dedicated to the development and validation of tools that support multicore software programming and FPGA acceleration, including Bambu for High-Level Synthesis and the XtratuM hypervisor with a level one boot loader for virtualization.

## I. Introduction

Space missions today face a growing need for on-board computing performance: low bandwidth communication links between spacecraft and Earth require sensor data to be pre-processed and compressed before transmission, and increasingly complex navigation algorithms cannot be operated remotely due to low latency constraints. The computational requirements for such tasks are rapidly approaching the limits of what space-grade processors and microcontrollers can offer, particularly because radiation-hardened components are inherently slower than general-purpose CPUs.

Instead, hybrid CPU-FPGA systems have drawn more and more attention [1]: they offer improved performances with acceptable overhead in size, power consumption, and cost (especially critical for CubeSat missions [2]), and they introduce the possibility of in-flight reconfiguration. Radiation-hardened FPGAs are thus a key enabling technology for space applications today, and the supply from European industries is quickly improving to meet current and future demand. European institutions such as ESA, CNES, and the EU funded several efforts to develop a new generation of rad-hard FPGAs, including projects such as BRAVE [10], VEGAS [11], OPERA [12], DAHLIA [1], and HERMES (this paper).

This paper presents HERMES - qualification of High pErformance pRogrammable Microprocessor and dEvelopment of Software ecosystem (https://www.hermes-h2020project.eu/), an H2020-funded project started in March 2021 in its intermediate stage. HERMES aims at validating and evaluating a state-of-the-art rad-hard FPGA according to the standards of the European Space Components Coordination (ESCC), and at integrating design and manufacturing technologies needed to deliver high reliability applications running on radiation-hardened integrated circuits.

The target platform for HERMES is the NG-ULTRA FPGA, designed in 28nm FD-SOI; NG-ULTRA is the world's first rad-hard SoC FPGA in 28nm, integrating a quad-core ARM R52 processor running at 600MHz (Figure 1). The platform will significantly improve the state of the art for rad-hard FPGAs thanks to a logic capacity of 550k LUTs running twice as fast as current rad-hard FPGAs with a power consumption four times smaller. The hardening techniques used in the design and manufacturing process for the NG-ULTRA board, alongside the FD-SOI technology, result in high reliability suitable to meet the requirements of critical aerospace applications, providing triple modular redundancy, error correction mechanisms, and memory integrity checks which are completely transparent to the application developer.

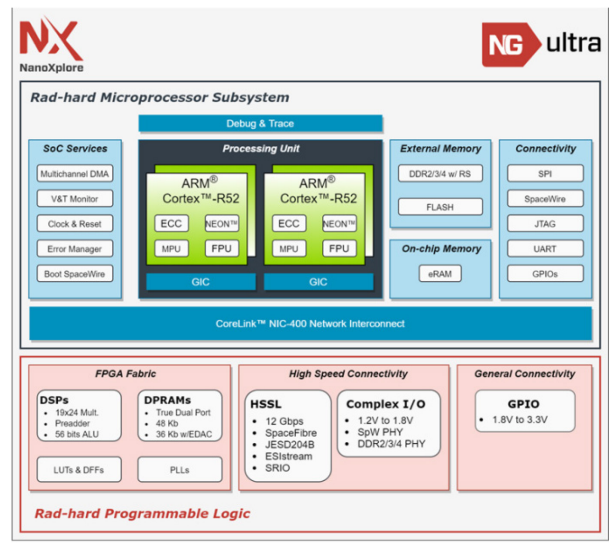

Fig. 1. NG-ULTRA architecture.

Besides finalizing the development and testing of the NanoXplore NG-ULTRA SoC, the project aims at providing a design ecosystem that will support the development of software applications and new soft IP blocks for FPGA. In particular, the existing NanoXplore synthesis toolkit will be extended to integrate the open-source Bambu High-Level Synthesis (HLS) tool [3], allowing fast prototyping of FPGA designs for space applications. Users will also receive support for platform virtualization through the XtratuM NextGeneration hypervisor [4] and a Generic Level 1 Boot loader.

The rest of the paper is structured as follows: we introduce High-Level Synthesis and the integration of Bambu into the

HERMES design flow in Section II, we describe the porting and the adaptation of the XtratuM Next Generation hypervisor in Section III, and the Generic Level 1 Boot loader in Section IV. We discuss the use cases that will be considered to evaluate the newly introduced tools in Section V and draw conclusions for the paper in Section VI.

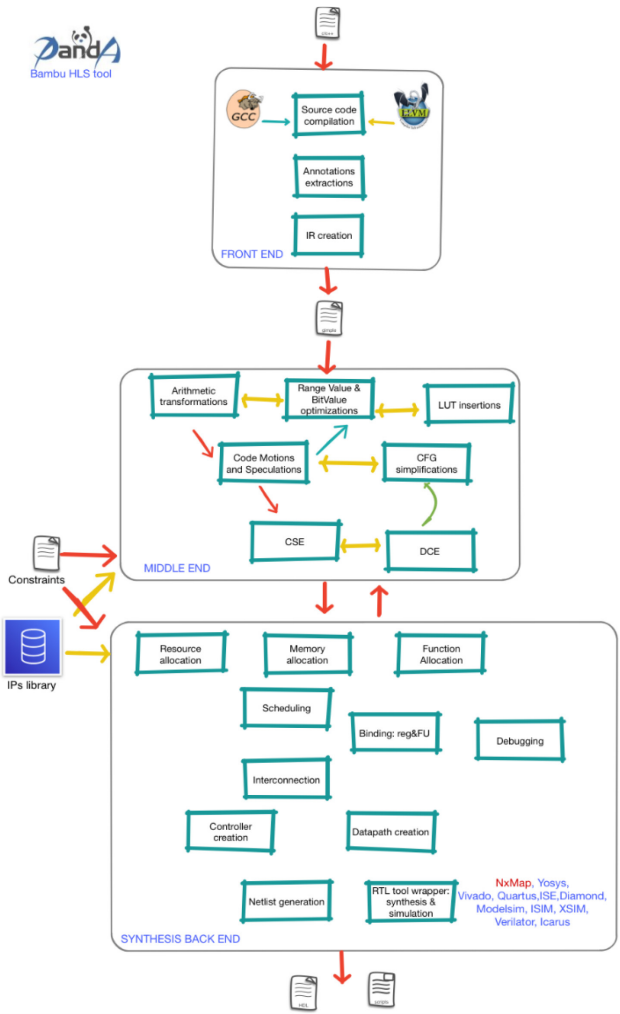

Fig. 2. Bambu HLS with its front-end, middle-end, and back-end steps.

## II. BAMBU HLS

HLS tools simplify the implementation of accelerators on FPGA by automating the most complex, time-consuming step in the development flow: instead of manually writing VHDL/Verilog code, the user only needs to provide a program written in a well-known software language such as C/C++, together with constraints about timing and resource utilization that the final design must satisfy. Generally, the High-Level Synthesis flow begins with a compilation step to analyze data dependencies and loops in the input C/C++ program, perform typical code optimizations, and generate a Control and Data Flow Graph (CDFG). Then three core steps are performed on the CDFG (resource allocation, scheduling, binding) to define the structure of the output hardware by assembling functional, storage, and communication units taken from a library of RTL components. Further optimization and analysis passes are applied in the front-end, middle-end, and back-end of the tool to generate efficient accelerator designs (Figure 2).

In the end, the generated HDL code is ready to be used in a commercial FPGA design tool for further analysis, logic synthesis, and deployment. In the past, the shorter development time offered by HLS used to be at odds with the efficiency of the generated designs. Instead, several commercial and open-source tools exist today [5,13] that can generate efficient designs, competitive in speed and resource utilization with hand-optimized RTL code.

The HERMES project selected Bambu HLS [3] to integrate capabilities to translate C code into Verilog/VHDL in its development ecosystem. In this way, the increased performance offered by FPGAs is made available also to software developers that do not have hardware design expertise. In HERMES, Bambu has been and will be extended to support new FPGA targets, architectural models, and input applications.

As a first step, the NanoXplore synthesis tool has been integrated with Bambu. NanoXplore provides the NXmap design suite (Figure 3) to transform user HDL RTL code into a bitstream for a specific NX device through logic synthesis, place, and route compilation steps; NXmap supports the complete NanoXplore radiation-hardened FPGA portfolio, and it includes both synthesis and static timing analysis tools. Seamless integration between Bambu and NXmap through the automatic generation of backend synthesis scripts raises the level of abstraction needed to design hardware accelerators.

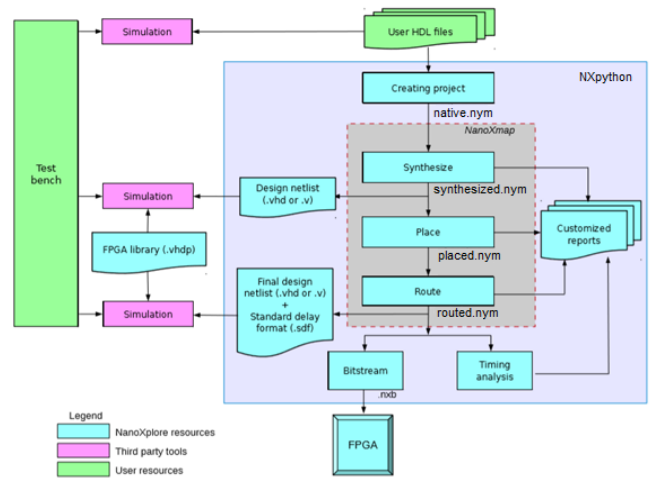

Fig. 3. NXmap design flow.

Since many of the optimizations applied by Bambu are target-aware, its back-end has been customized to support the NG-ULTRA FPGA through a pre-characterization and performance estimation process. In fact, all library components used during the HLS flow need to be annotated with information such as resource occupation and latency under different clock period constraints (FPGAs do not have a fixed operating frequency, this can be decided by the designer or forced by the attached devices such as sensors or actuators). Performance estimation of library components is essential to perform aggressive optimizations, so Bambu integrates a characterization tool called Eucalyptus to synthesize different configurations of library components and collect the resulting latency and resource consumption metrics as XML files in the Bambu library. The configurations are obtained by specializing a generic template of the resource component (e.g., a multiplier or an adder) according to the bit widths of its input and output arguments, and to the number of pipeline stages. Between the integration of the logic synthesis backend based on NXmap and the characterization run through Eucalyptus it was also necessary to correctly map

Bambu library components on the actual DSPs and True Dual Port RAMs available on the NG-ULTRA fabric: since the mapping happens through behavioral HDL templates, the components used by Bambu for arithmetic operations and the storage modules have been customized to be compliant with the NXmap synthesis guidelines.

The integrated ARM processor on the NG-ULTRA board uses the AXI4 protocol [7] interfaces to communicate with the rest of the system; therefore, support for AXI4 interfaces has been added to Bambu to facilitate the integration of HLS-generated accelerators with the host processor. This allows the user to automatically generate the necessary AXI4 master interfaces and modules controlling the AXI signals, with no protocol knowledge required. Bambu can handle multiple interfaces associated with the same module, and different parameters can share the same interface; data accesses are automatically mapped to the appropriate AXI controller. In addition to the module generation, Bambu supports the creation of a testbench that includes the AXI4 slave counterparts of the master interfaces, so that data exchange can be simulated to verify its correctness. Memory delay estimates can also be configured to assess the performance of the application considering also data transfers. The generated interface code is fully functional and supports unaligned memory accesses, and it is currently only available in Verilog. Additional development efforts will be dedicated to improving performance, especially for applications that need to access considerable amounts of data: for example, adding support for prefetching and caching mechanisms might drastically reduce the average access time. Furthermore, Bambu will be extended to support the customization of cache sizes, associativity, and other features that will help the user generate HDL code tailored to a specific application use case with minimal effort.

The HERMES project use cases include applications based on artificial intelligence, which might contain multiple parallel execution flows (i.e., coarse-grained parallelism); when synthesized through an HLS tool, the complexity of the finite state machine controllers for such applications grows exponentially, leading to considerable resource consumption and latency overheads. To solve this problem, Bambu has been extended to efficiently synthesize dynamically controlled accelerators and integrated in a compiler-based toolchain that facilitates the extraction of coarse-grained tasks and data dependencies from an input machine learning application designed and trained in a high-level programming framework [14]. Future developments will focus on other optimization techniques and architectural templates that can answer the specific needs of artificial intelligence algorithms and the requirements of aerospace applications.

## III. XtratuM NextGeneration

Recent evolutions of the hardware technology used in embedded real-time systems have resulted in very powerful, reliable, and cost-effective multi-core platforms. This trend has increased interest in software architectures with multiple applications sharing a multi-core platform as the best solution for embedded real-time applications. The growing need to reduce the size, weight, power consumption, and cost in various markets such as aerospace, automotive, or IoT has further boosted the importance of multi-core platforms. Sharing a common hardware platform requires implementing mechanisms to ensure that applications do not interfere with each other; virtualization solutions play an essential role in achieving this thanks to their time and space partitioning (TSP) concept.

Virtualization is an operating system paradigm in which a kernel allows multiple isolated operating systems hosted on a single physical computer system. These virtual computer systems are often referred to as virtual machines or partitions.

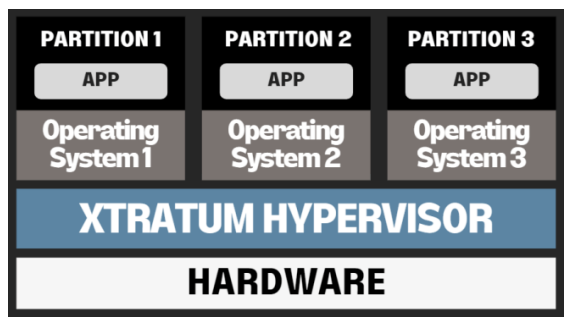

Fig. 4. Partition diagram.

Virtualization plays a vital role in the context of the NG-ULTRA software ecosystem, and specifically in HERMES, since it is a critical element to fully exploit the quad-core ARM R52 integrated on the platform.

The tool considered for virtualization in HERMES is the XtratuM hypervisor [6]. XtratuM is a bare-metal space-qualified hypervisor aimed at safe and efficient execution of embedded real-time systems. It enables applications to share the same (multicore) hardware platform allowing easier reuse of legacy certified applications, more straightforward dynamic software updates, and size, weight, power, and cost reduction for safety-critical systems.

In the HERMES project, the XtratuM hypervisor will implement partial virtualization, where the hypervisor provides partitions with a similar interface to the one of the underlying hardware platform (i.e., the real/physical machine). Such a partition is thus capable of hosting generic software with minor modifications. Currently, the implementation phase is considered finished as the XtratuM hypervisor has been successfully adapted to the NG-ULTRA SoC-based board, giving support to the four cores provided by the board, thus enabling parallel computing. The next step in the project is to start the qualification phase following the ECSS at DAL-B.

## IV. Generic Level 1 Boot Loader Development

Any hardware platform needs to be configured at start-up before being able to run high-level software. This configuration is performed through a boot-loading sequence, where each step is a piece of software that performs the necessary configuration of the processing system, preparing it to execute higher-level software. At the end of the sequence, the boot loader loads the upper-level software (bare metal application or operating system) and gives it control of the platform.

In the embedded domain, boot loaders are usually part of a Board Support Package and they must prove compliance with additional specific requirements of availability, reliability, safety, and security. Moreover, they are generally smaller in size and complexity, sacrificing versatility for the sake of simplicity. For the space domain, such constraints translate at the software level into the application of development processes following the ECSS standard, i.e., ECSS-Q-ST-80C and ECSS-Q-ST-40 standards. Interested readers more familiar with other domains applying ISO 26262, IEC 61508 or DO-178 can refer to [8] for some insights on ECSS standards.

The NG-Ultra System on Chip (SoC) boot procedure is split into the following main stages:

- The first boot-loading step is BL0 (for Bootloader level 0). It is a small application hard-coded into the SoC internal ROM that fetches a binary executable (called BL1 for Bootloader level 1) from either local boot FLASH memory or remotely from the SpaceWire bus.
- BL1 then initializes the main hardware components, loads the eFPGA matrix configuration (i.e., the bitstream), and loads and branches to a later applicative stage.
- An additional BL2 stage or the final application-dependent software finalizes the hardware configuration and can deploy itself on all the available processor cores.

On the NanoXplore platforms, BL0 was developed in the scope of the DAHLIA H2020 project [1] and it is hardcoded in the SoC eROM; BL1 instead is developed in the scope of the HERMES project and is field loadable. HERMES inherited a generic BL1 specification and hardware components drivers (such as DDR controller and flash controller) that were jointly developed by AIRBUS Defence and Space and THALES Alenia Space in the scope of CNES R&T study R-S19/BS-0004-061, an activity that was carried out in close collaboration with NanoXplore for integration with their SDK/BSP package.

One of the main principles followed in the BL1 specification and design has been to keep things simple and focus only on functions deemed mandatory and common to any mission scope. BL1 is expected to be either reused as is to branch to a custom BL2 or Application Software, or customized and modified according to mission needs by the final customer. The common functionalities of the BL1 for the NG-ULTRA SoC include the following:

- Initialization of the master CPU#0 registers, caches, and exceptions at the most privileged level.
- Initialization of mandatory hardware resources such as Clock PLLs, DDR controller, Flash controller, SpaceWire controller, and Tightly Coupled Memories.
- Initialization of Memory Protection Unit allowing access to local Tightly Coupled Memories, embedded RAM, and external DDR.
- Management of a load list, either stored in Flash or remotely received from SpaceWire following a custom protocol, describing a set of application software to be deployed to memory, and bitstream to be programmed in the eFPGA matrix.
- Management of integrity of deployed software and proper eFPGA matrix programming.
- Management of basic redundancy for software components stored in Flash (either through TMR or through sequential accesses to multiple hardware Flash components).
- Generation of a BL1 boot report made available for next-stage software.

An example of a power-up sequence from boot flash covering BL0 to BL2 stages is illustrated in Fig. 5.

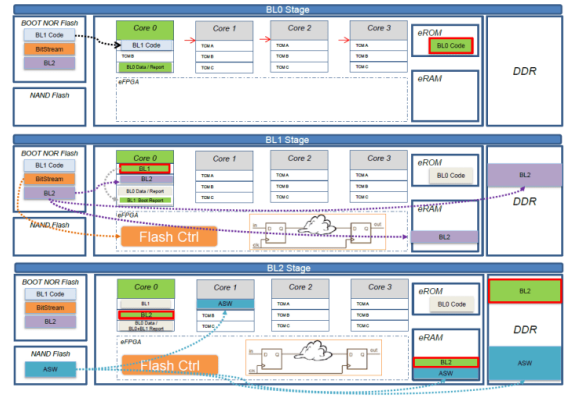

Fig. 5. BL1 step part of the complete NG-ULTRA boot sequence.

A comprehensive qualification datapack will be generated during the HERMES project composed of a consolidated version of mandatory documents paving the road toward ECSS level B qualification (SRS, SUITP/SUITR, SVTS, SValP/SValR, and SUM). The deliverable will contain the BL1 source code and testing tool suite implementation, covering unitary, integration, and validation source code using open-source software tools (*gcc* compiler, *gcov* for coverage, google test suite). Currently, there is only one exception to the open-source approach, which is the TRACE32 hardware-based debugger tool from Lauterbach. At time of writing, the definition of the specification for the boot loader level 1, the design and implementation of the software, and the unit tests have been completed. Initial source code delivery and specification validation have recently started, the following steps will cover the finalization of the validation and the final delivery of the data pack.

V. USE CASES

HERMES brings together a group of unique, highly relevant, and qualified partners already working together in previous and current EU or ESA activities, and the HERMES timetable and work packages have been coordinated with the NG-ULTRA development activities to limit project execution risks. HERMES will directly impact the European space end-users by giving them access to both the flight model of the most advanced radiation-hardened programmable SoC and to an optimized software ecosystem. The ESCC has evaluated NG-ULTRA as a key step for future European technology independence, and HERMES fully complements the ongoing activities of other ESA and CNES projects.

The outcomes of the HERMES project will be validated through the execution of representative space use cases. This will allow to illustrate a set of key benefits and measurable metrics essential for the exploitability of tools developed by the project, and it will prove the suitability of the developed software ecosystem in the specific context of space applications.

Regarding the evaluation of the HLS component, the application use cases cover image and vision processing algorithms, software-defined algorithms, and artificial intelligence applications. The evaluation will consist of generating IP cores from the source code of the applications through Bambu, and of the IP integration and execution on a representative NG-ULTRA platform. Metrics regarding both the functionality and usability of the HLS tool and the performance of the generated IP core will be collected and evaluated.

Regarding the evaluation of XtratuM hypervisor, a use case inherited from the SELENE H2020 [9] project will be adapted to test the virtualization tools. The application, in this case, includes representative elements of space mission control such as an Attitude and Orbit Control system (AOCS), Visual Based Navigation image processing, Electrical Orbit Raising algorithms.

## VI. Conclusion

The ambition of the HERMES project is to offer Europe new advanced radiation-hardened SoC FPGA platforms for space applications, and for additional markets that share similar harsh environments constraints and high-reliability requirements. (For example, avionics, automotive, and energy production also need extended temperature range, extended lifetime, and fault management). Indeed, FPGAs are very versatile and they can serve many different applications outside a specific market. Integrating an HLS tool and a virtualization hypervisor in an integrated software environment will improve productivity for developers, that will have access to the NG-ULTRA platform together with its software ecosystem. More specifically, the use of High-Level Synthesis through Bambu will raise the level of abstraction needed to develop critical applications on rad-hard FPGAs. The XtratuM Next Generation hypervisor and qualification activities carried out by the project will achieve the ECSS level B required by onboard software in space applications.

## Acknowledgment

The HERMES project has received funding from the European Union's Horizon 2020 research and innovation Programme under grant agreement No 101004203.

## References

[1] J.-L. Poupat et al. DAHLIA: Very High Performance Microprocessor for Space Applications. Presented at Data Systems in Aerospace (DASIA 2018).

[2] A. D. George and C. M. Wilson, "Onboard Processing With Hybrid and Reconfigurable Computing on Small Satellites," in Proceedings of the IEEE, vol. 106, no. 3, pp. 458-470, March 2018.

[3] F. Ferrandi, V. G. Castellana, S. Curzel, P. Fezzardi, M. Fiorito, M. Lattuada, M. Minutoli, C. Pilato, and A. Tumeo, "Bambu: an open-source research framework for the high-level synthesis of complex applications," in DAC 2021: 58th ACM/IEEE Design Automation Conference, 2021, pp. 1327–1330.

[4] E. Carrascosa, J. Coronel, M. Masmano, P. Balbastre, and A. Crespo. 2014. stratum hypervisor redesign for LEON4 multicore processor. SIGBED Rev. 11, 2 (June 2014), 27–31.

[5] R. Nane et al., "A Survey and Evaluation of FPGA High-Level Synthesis Tools," in IEEE Transactions on Computer-Aided Design of Integrated Circuits and Systems, vol. 35, no. 10, pp. 1591-1604, Oct. 2016.

[6] Fentiss. XtratuM hypervisor. 2022. https://fentiss.com/products/hypervisor/.

[7] AMBA® AXI™ and ACE™ Protocol Specification. https://developer.arm.com/documentation/ihi0022/e/ .

[8] Joseph Machrouh, Jean-Paul Blanquart, Philippe Baufreton, Jean-Louis Boulanger, Hervé Delseny, et al.. A cross-domain comparison of software development assurance standards. Embedded Real-Time Software and Systems (ERTS2012), Feb 2012, Toulouse, France.

[9] SELENE H2020 project, "Highly integrated Satellite Control and data management," https://www.selene-project.eu/highly-integrated-satellite-control-and-data-management/.

[10] "High Density European Rad-Hard SRAM-Based FPGA - First Validated Prototypes – BRAVE" https://www.esa.int/Enabling_Support/Space_Engineering_Technology/Shaping_the_Future/High_Density_European_Rad-Hard_SRAM-Based_FPGA-_First_Validated_Prototypes_BRAVE

[11] "Validation of European high capacity rad-hard FPGA and software tools (VEGAS)", supported by H2020 under grant agreement n. 687220.

[12] "Space Qualification and Validation of High Performance European Rad-Hard FPGA (OPERA)", supported by H2020 under grant agreement n. 821969.

[13] J. Cong, J. Lau, G. Liu, S. Neuendorffer, et al., "FPGA HLS Today: Successes, Challenges, and Opportunities," ACM Transactions on Reconfigurable Technology and Systems, vol. 15, no. 4, pp. 1–42, 2022.

[14] S. Curzel et al., "End-to-End Synthesis of Dynamically Controlled Machine Learning Accelerators," in IEEE Transactions on Computers, vol. 71, no. 12, pp. 3074-3087, 1 Dec. 2022, doi: 10.1109/TC.2022.3211430.